\documentclass[%
 reprint,
superscriptaddress,
nofootinbib,
 amsmath,amssymb,
 aps,
]{revtex4-1}

\usepackage{graphicx}
\usepackage{dcolumn}
\usepackage{bm}
\usepackage{amsmath,amsfonts,amssymb,slashed,upgreek,color}
\usepackage{multirow}
\usepackage{xcolor}
\usepackage{hyperref}
\usepackage[export]{adjustbox}
\usepackage{ulem}

\graphicspath{{plots/}}

\newcommand\ee{\end{equation}}
\newcommand\be{\begin{equation}}
\newcommand{\HH}{\mathcal H}

\newcolumntype{C}[1]{>{\centering\arraybackslash}p{#1}}
\newcolumntype{L}[1]{>{\raggedright\arraybackslash}p{#1}}
\newcolumntype{R}[1]{>{\raggedleft\arraybackslash}p{#1}}
\definecolor{bittersweet}{rgb}{1.0, 0.44, 0.37}

\begin{document}

\hyphenation{Stuckelberg}


\title{A new non-linear instability for scalar fields}

\author{Farbod Hassani}
\email{farbod.hassani@astro.uio.no}
\affiliation{%
Institute of Theoretical Astrophysics, University of Oslo, 0315 Oslo, Norway
}%
\affiliation{%
Universit\'e de Gen\`eve, D\'epartement de Physique Th\'eorique and Centre for Astroparticle Physics, 24 quai Ernest-Ansermet, CH-1211 Gen\`eve 4, Switzerland
}%
\author{Pan Shi}
\affiliation{%
Universit\'e de Gen\`eve, D\'epartement de Physique Th\'eorique and Centre for Astroparticle Physics, 24 quai Ernest-Ansermet, CH-1211 Gen\`eve 4, Switzerland
}%
\author{Julian Adamek}
\affiliation{%
Universit\"at Z\"urich, Institute for Computational Science, Winterthurerstr.\ 190, CH-8057 Z\"urich, Switzerland
}%
\author{Martin Kunz}
\affiliation{%
Universit\'e de Gen\`eve, D\'epartement de Physique Th\'eorique and Centre for Astroparticle Physics, 24 quai Ernest-Ansermet, CH-1211 Gen\`eve 4, Switzerland
}%
\author{Peter Wittwer}
\affiliation{%
Universit\'e de Gen\`eve, D\'epartement de Physique Th\'eorique and Centre for Astroparticle Physics, 24 quai Ernest-Ansermet, CH-1211 Gen\`eve 4, Switzerland
}%

\date{\today}

\begin{abstract}
In this letter we introduce the non-linear partial differential equation (PDE) $\partial^2_{\tau} \pi \propto (\vec\nabla \pi)^2$ showing a new type of instability.
Such equations appear in the effective field theory (EFT) of dark energy for {the} $k$-essence model as well as in many other theories  based on the EFT formalism. 
We demonstrate the occurrence of instability in the cosmological context using {a} relativistic $N$-body code, and we study it mathematically in 3+1 dimensions within spherical symmetry. We show that this term dominates for the low speed of sound limit where some important linear terms 
are suppressed.
\end{abstract}

\pacs{Valid PACS appear here}
\maketitle


\section*{Introduction}
One of the main goals of the upcoming large cosmological surveys \cite{Amendola:2016saw, 2015aska.confE..19S,4MOST:2019,Aghamousa:2016zmz} is to understand the physical mechanism behind the mysterious late-time accelerating expansion of the Universe \cite{2016A&A...594A..13P, Scolnic:2017caz, Alam:2016hwk}. Accurate modelling of the current viable dark energy {(DE) and} modified gravity {(MG)} candidates over all scales of interest is critical for the highly precise data sets that these surveys will deliver over the coming decade. 

To study many possible models that include a {DE} component, or where the theory of gravity is altered, the EFT 
framework has been suggested \cite{Pich:1998xt, Gubitosi:2012hu, Frusciante:2017nfr, Creminelli:2008wc}. In the EFT scheme a general form of the action is considered up to a certain energy 
scale and the idea is that only some degrees of freedom are relevant below that 
scale, while those degrees of freedom that describe properties of system at higher energy scales can be integrated out \cite{HARTMANN2001267,Hassani:2020agf}. The EFT of DE is particularly useful for cosmologists as one can map most of the interesting MG/DE theories to this language by choosing the set of free parameters appropriately. The EFT of DE thus provides a 
framework for a generic study of DE/MG theories \cite{Gleyzes:2013ooa, Gleyzes:2014rba, Bellini:2014fua}.

As a first step toward implementing the EFT of DE in an $N$-body simulation, we have developed the $k$-evolution code \cite{kevolution} based on \textit{gevolution} \citep{Adamek:2016zes, Adamek:2015eda}. $k$-evolution is able to simulate non-linear structure formation with $k$-essence dark energy \cite{ArmendarizPicon:2000dh,ArmendarizPicon:2000ah}. 

Our extensive numerical studies using $k$-evolution have led us to the discovery of a new type of non-linear instability that appears naturally in such EFT expansions and is not limited to the $k$-essence type of theories. This instability is not in the form of rapid growth of the scalar field but rather is an instability in the mathematical sense, in which the scalar field {solution} ceases to exist at a finite ``blow-up'' time, leading to the breakdown of the EFT framework. 

\section*{EFT equations of motion for $k$-essence \label{Field_equations}} 
The action for a general scalar field theory constructed from the scalar field $\phi$ and the kinetic term $X \doteq g^{\mu \nu} \partial_\mu \phi \partial_\nu \phi $ can be written as
 \begin{equation}
S 
= \int  d^4 x \sqrt{-g} P (X, \phi)  \, , \label{eq:kessence_action}
 \end{equation}
where $P$ is in general an arbitrary scalar function of its arguments, $g$ is the determinant of the metric and the integral is taken over the four-dimensional space-time. This class of theories is known as $k$-essence  \cite{ArmendarizPicon:2000dh,ArmendarizPicon:2000ah}.
In the EFT of DE framework, assuming small scalar field fluctuations, a 3+1 split of space-time can be defined by using the scalar field as a `clock' to define constant-time hypersurfaces. Writing the action as an expansion in terms of geometric scalars we obtain \cite{Cheung:2007st,Creminelli:2008wc},
\begin{align}
\label{EFTaction}
S =\int d^4 x \sqrt{-g} \Big [ \frac{M_\text{pl}^2}{2} R -\Lambda (t) & -c(t) g^{00}  +\frac{M_2^4(t)}{2} \left (\delta g^{00}  \right )^2  \Big]
\end{align}
where ${M_\text{pl}}$ is the Planck mass, $R$ is the four-dimensional Ricci scalar, $\Lambda(t)$, $c(t)$, and $M_2^4(t)$ are time-dependent functions and $\delta g^{00}$ is the perturbation of $g^{00}$ around its background value. We  have ignored terms that are of higher order in the fluctuations $\delta g^{00}$, because these terms are negligible in the weak-field expansion relevant for cosmology. The scalar field and its perturbation $\pi$ can be reintroduced, as usual in this framework, with the St\"uckelberg trick. See \cite{Hassani:2019lmy} for more details.

The variation of the action with respect to the metric results in the gravitational field equations \cite{Adamek:2016zes}, while the variation 
with respect to the scalar field perturbation $\pi$ results in a non-linear PDE for the $k$-essence scalar field,  
\begin{align}
& \partial_{\tau}^2\pi  + \HH (1-3w) \partial_{\tau}\pi  + \Big( \partial_{\tau} \HH - 3w \HH^2 + 3 c_{s}^{2} (\mathcal{H}^{2} - \partial_{\tau} \HH)  \Big)  \pi   \nonumber \\ & 
-  \partial_{\tau}\Psi +3\HH (w- c_s^2) \Psi - 3 c_{s}^{2} \partial_{\tau}\Phi
-c_{s}^{2}  \nabla^2\pi 
\nonumber \\ &
= {\mathcal{N}(\pi,  	\partial_\tau \pi,  \vec{\nabla} \pi,    \vec{\nabla} \partial_{\tau} \pi,\nabla^2 \pi) } \,, \label{full_1Deq}
\end{align}
where $ \mathcal{N}(\pi,  	\partial_\tau \pi,  \vec{\nabla} \pi,    \vec{\nabla} \partial_{\tau} \pi,\nabla^2 \pi)$ includes all the non-linear terms in the equation,
\begin{align}
&  \mathcal{N}(\pi,  	\partial_\tau \pi,  \vec{\nabla} \pi,    \vec{\nabla} \partial_{\tau} \pi,\nabla^2 \pi) 
 =    - \frac{\HH}{2}  \big(5c_s^2  + 3w  -2\big) \; \boxed{{(\vec{\nabla} \pi)^2}}
 	\nonumber \\ &
 	+ 2(1- c_s^2)  \vec{\nabla} \pi \cdot \vec{\nabla}\partial_{\tau}  \pi
	 - \Big[ (c_s^2-1) \big( \partial_{\tau}  \pi + \HH \pi -\Psi \big)
	 \nonumber \\ &
	 + c_s^2 (\Phi - \Psi) + 3 \HH c_s^2 (1+w)\pi  \Big] \nabla^2 \pi
	 + (2 c_s^2 -1)  \vec{\nabla} \Psi  \cdot \vec{\nabla} \pi   	  
	 \nonumber \\ & 
	  - c_s^2 \vec{\nabla} \Phi \cdot \vec{\nabla} \pi 
 +\frac{3(c_s^2 -1) }{2}  \partial_i \Big(\partial_i \pi (\vec{\nabla} \pi)^2\Big) \,.
\label{terms_non_linear}
\end{align}
 We have parametrized the {model} with {an} equation of state $w$ and {a} speed of sound  $c_s$ which, respectively, relate
 to the EFT parameters through
\be 
w= \frac{c - \Lambda}{c + \Lambda}\,, \;\;  c_{s}^{2} = \frac{c}{c+2 M_{2}^{4}}\,.
\ee
In these expressions, $\HH>0$ is the (conformal) Hubble parameter that gives the expansion rate of the Universe and $\Phi$, $\Psi$ are the two gravitational potentials in longitudinal gauge. The equation of state parameter $w$ is close to $-1$ if the $k$-essence field is to play the role of the dark energy.

 The equation {of motion} as well as the stress energy tensor are discussed in detail in \cite{Hassani:2019lmy}. However, in this article we only focus on the PDE and the instability which is 
 caused by the first term on the right-hand side of Eq.\ \eqref{terms_non_linear}. 
\section*{The New instability} \label{New instability}
Numerical simulations using $k$-evolution 
empirically show
that the evolution under Eq.\ \eqref{full_1Deq} is unstable for 
small values of the speed of sound $c_s$. We also find that
there is a 
critical value
$c_s^*$ 
such that for a speed of sound
lower than $c_s^*$ the evolution becomes singular 
at a finite time, before 
the Universe reaches its present age. On the other hand, for speeds of sound much larger than the critical value, the singularity is avoided altogether. The
instability forms in the regions with the highest curvature of the gravitational potential (center of halos) and the blow-up time depends on the initial curvature of the potential wells. 

\begin{figure*}
\begin{minipage}{\textwidth}
\vspace*{1cm}
 {\hspace*{-1.5cm}{\includegraphics[scale=0.22, trim=-1cm 3cm 0 0cm]{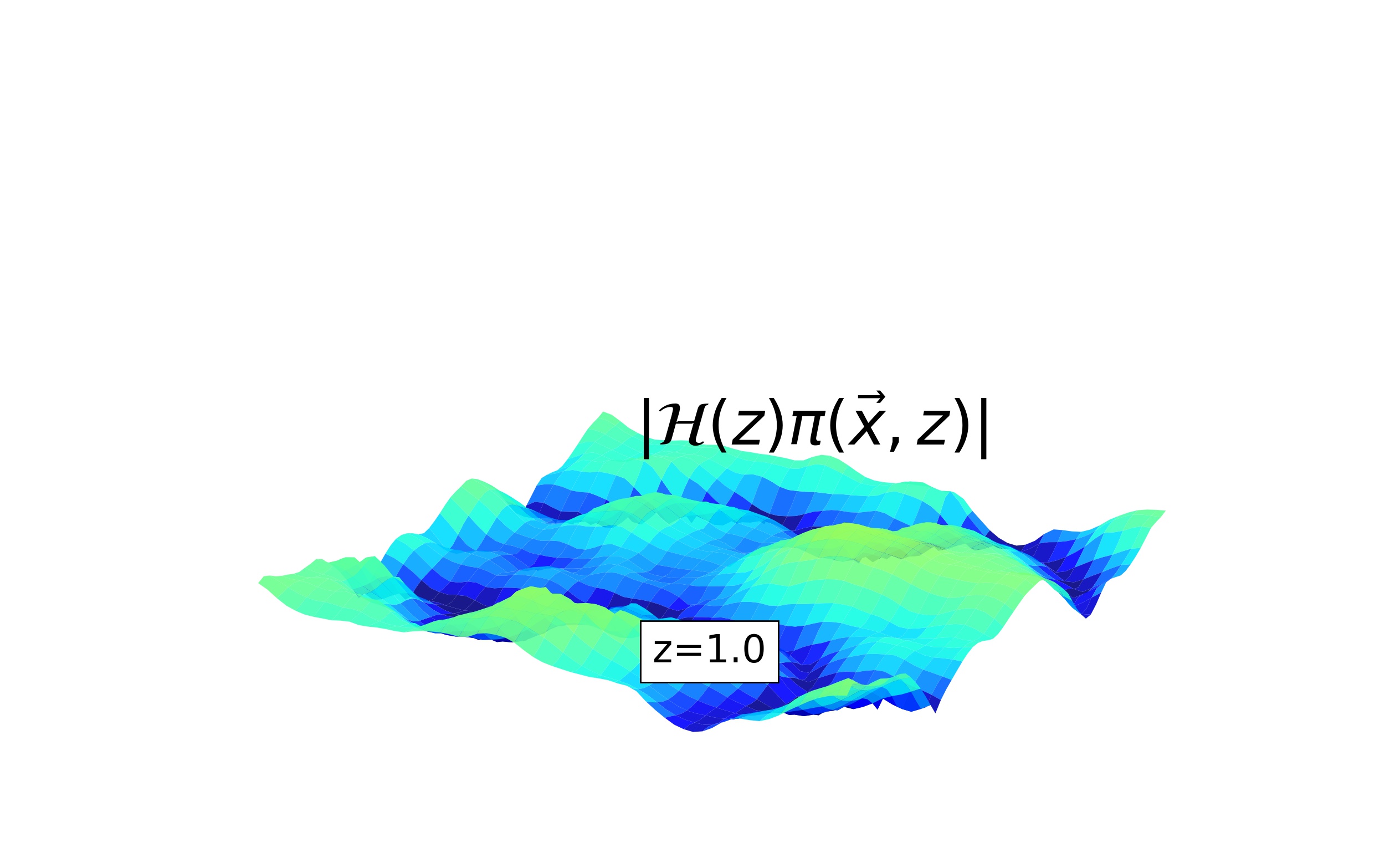} }}
 {\hspace*{-1.5cm}{\includegraphics[scale=0.22, trim=-1cm 3.5cm 0 0cm]{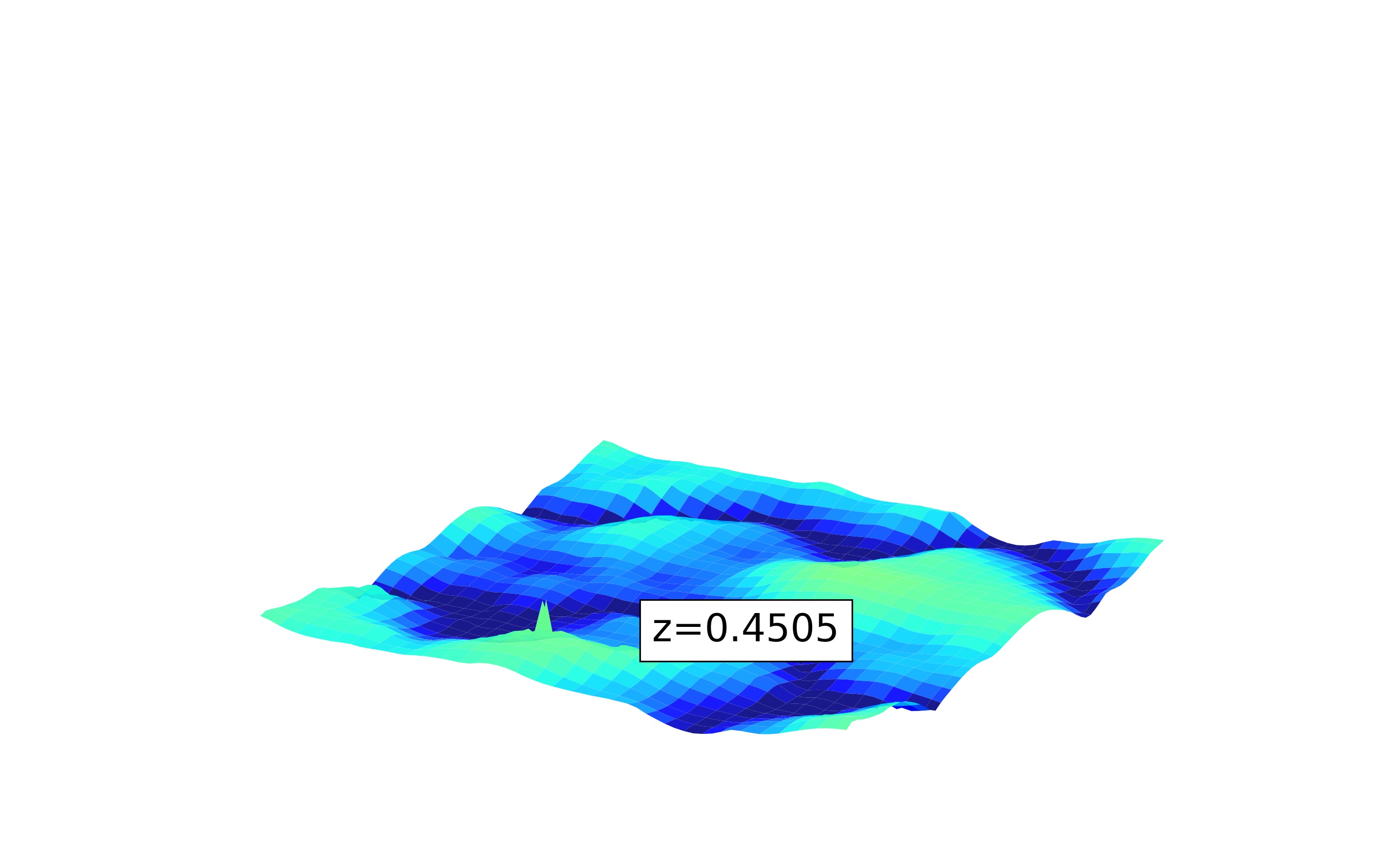} }}
  {\hspace*{-1.5cm}{\includegraphics[scale=0.22, trim=-1cm 4.0cm 0 -1cm]{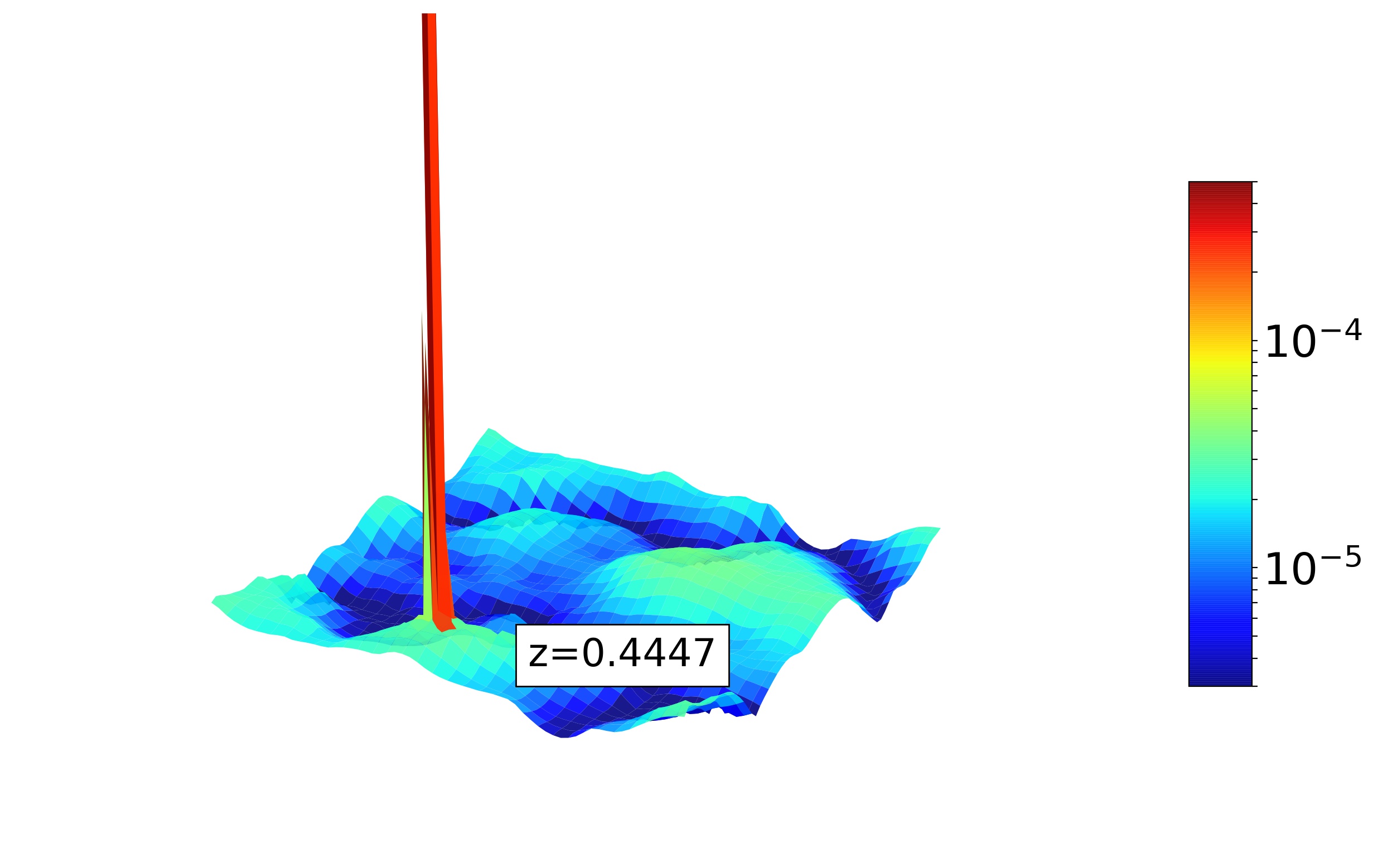} }}\hfill
 \caption{
 From left to right: The evolution of the absolute value of the scalar field perturbation times the Hubble parameter, $\mathcal{H} \pi$, from $k$-evolution in time is shown. At the time corresponding to $z \approx 0.45$ the system blows up due to the non-linear instability. This instability acts very quickly and leads to a divergence of $\pi$ in finite time. Note that for the sake of illustration we plot the absolute value of the scalar field and 
 that the blow-up occurs at a local minimum.}
 \label{1D_snapshot}
 \end{minipage}
 \end{figure*}

 In Fig.\ \ref{1D_snapshot}, we show the evolution of the scalar field perturbation $\pi$ times the Hubble parameter $\mathcal{H}$ on a 
 2D slice taken
 from a cosmological simulation. As we can see the instability is formed suddenly at a certain redshift and the scalar field solution ceases to exist at this time.
 
By considering subsets of the terms in $\mathcal{N}$, we find that the non-linear instability is generated 
by the term quadratic in the gradient,
$(\vec{\nabla}\pi)^2$ in Eq.\ \eqref{full_1Deq}. 
However, for large speed of sound, Eq.\ \eqref{full_1Deq} is dominated by the linear terms rather than the non-linear ones. This is expected because in this limit 
the sound horizon of the scalar field is of order Hubble scale 
and the scalar field perturbations decay inside the sound horizon. Thus we do not expect to have non-linear effects at small scales. On the other hand, in the low speed of sound limit the corresponding sound horizon is small and in Eq.\ \eqref{full_1Deq} 
the linear restoring force, $c_s^2 \vec{\nabla}^2 \pi$, is suppressed.
As a result  the non-linear terms become important and the instability forms.
   
\section*{Mathematical results}  \label{Mathematics}
In this section we are going to mathematically show that the 
problematic
non-linear term in Eq.\ \eqref{terms_non_linear}
will inevitably give rise to a singularity in finite time once it dominates the dynamics. 
 We only focus on the single problematic term rather than the whole PDE. In other words, we consider the simple non-linear equation in 3+1 dimensions
\be
\partial_{\tau}^2 \pi  = \alpha  \vec{\nabla} \pi \cdot\vec{\nabla} \pi \, , \label{3D_PDE}
\ee
where $\alpha$ is the coefficient of the problematic term\footnote{In a general EFT of DE theory $\alpha$ 
depends on
the EFT parameters. For the EFT of $k$-essence we have $\alpha =- \frac{\HH}{2}  \big(5c_s^2  + 3w  -2\big) $.}.
Around the local extrema $x_*$ where $\vec{\nabla} \pi|_{x_*}=0$ we choose spherical coordinates to study the behaviour
of the solution. This is a reasonable choice as according to Eq.\ (\ref{3D_PDE}) such a point remains an extremum at all times\footnote{In addition to $\vec{\nabla} \pi|_{x_*}=0$ we also need to have $\vec{\nabla} \frac{\partial\pi}{\partial \tau}|_{x_*}=0$ at initial time, which is a reasonable assumption. In our numerical studies we consider the scalar field and its time derivative to be zero at initial time and the scalar field being generated solely by the gravitational coupling to the matter perturbations.}. Moreover, the spherical symmetry is preserved under time evolution. Thus, for such points and for spherically symmetric initial conditions we have the following PDE,
 \be
\partial_{\tau}^2 \pi (\tau,r) = \alpha \left[\partial_{r} \pi (\tau,r)\right]^2\,, \label{1D_PDE}
\ee
This PDE is unstable independently of the sign of $\alpha$: in the case of positive $\alpha$ the 
singularity occurs at local minima of the scalar field whereas for negative $\alpha$ it occurs at local maxima.
For the 
EFT of $k$-essence $\alpha>0$ and the instability is formed in the minima which 
generally coincide with the centers of halos.

We may choose units such that $\alpha = 1$ and we solve this equation for the initial conditions $\pi(\tau_0 , r)$ and $ \partial_{\tau} \pi(\tau_0 , r)$. {It is worth noting that} even if we assume $\pi(\tau_0 , r) = 0$ and $\partial_{\tau}\pi(\tau_0 , r) = 0$  in a cosmological scheme the gravitational potential $\Psi$ would eventually source the scalar field as is evident from Eq.\ \eqref{full_1Deq}. Here we instead consider a general initial condition for the scalar field and do not keep the gravitational source term.
One particular solution to the non-linear PDE \eqref{1D_PDE} is given by
\be
 \pi_s (\tau,r)= \kappa(\tau) r^2\,, 
 \label{eq:rsquared}
 \ee
where $2 \kappa(\tau) $ represents the curvature of the scalar field  $\pi_s (\tau,r)$ in time, and $\kappa(\tau)$ is a solution to the ordinary differential equation  
\be 
\partial_{\tau}^2{\kappa}( \tau) = 4  \left[\kappa (\tau)\right] ^2\,. \label{kappa_eq}
\ee

The initial conditions  $\kappa(\tau_0) $ and $\partial_{\tau} {\kappa}(\tau_0)$ 
{can be}
obtained based on the assumed initial condition for $ \pi (\tau_0,r)$ and $\partial_{\tau} \pi (\tau_0,r)$.
We can think of this ODE as Newton's second law with the force $F(x) = 4 x^2 $, which corresponds to a potential $V(x) = - \frac43 x^3$. No matter what the initial conditions for $x(0)$ and $\frac{dx}{d\tau}(0)$  or equivalently $\kappa(0)$ and $\frac{d\kappa}{d\tau}(0)$ (except  $\kappa(0) = \frac{d\kappa}{d\tau}(0)=0$ ) are, a particle on this potential rolls to $+\infty$ eventually. Here we are going to show that in fact the particle (in our case the curvature of the scalar field) goes to infinity in a finite time $\tau_b$. To solve Eq.~\eqref{kappa_eq} we multiply both sides by $\kappa'=\frac{d \kappa}{d\tau}$,
\begin{align}
& \frac12 \frac{d (\kappa'(\tau)^2)}{d \tau} = \frac{4}{3} \frac{d (\kappa (\tau)^3)}{d\tau} \, .
\end{align}
Integrating results in the following expression,
\begin{align}
&\kappa'(\tau)^2 =\kappa' (0)^2 + \frac{8}3 \kappa(\tau)^3 - \frac{8}3 \kappa(0)^3 \, .
\end{align}
Integrating once more 
we obtain,
\be
\int_{\kappa(0)}^{\kappa(\tau)} \frac{d \kappa}{\sqrt{\kappa' (0)^2 + \frac{8}3 \kappa^3 - \frac{8}3 \kappa(0)^3 } } = \int_0^{\tau} d\tau' = \tau \, .
\ee
Changing the integration variable from $\kappa$ to $s$ for $s^3 = \frac{8}{3} \frac{\kappa^3 }{C}$ and $C =\kappa' (0)^2 - \frac{8}{3}\kappa(0)^3$ we find that $\tau$ is bounded by
\be
\tau_b = \left(\frac{3 }{8}\right)^{\frac13} \left(\frac{1}{ C} \right)^{\frac16} \int_{s(\kappa_0)}^{\infty} \frac{d s}{\sqrt{1+s^3}} \, , \label{eq:blowup}
\ee
i.e.\ the solution blows up in finite time.
In the cosmological context we can set $\kappa(0)=0$ so that $C>0$ and $s(\kappa_0)=0$, giving us a blow-up time of
\be
\tau_b =  \left(\frac38\right)^{\frac 13} \left(\frac{1}{\kappa' (0)}\right)^{\frac13}  \frac{2 \Gamma [\frac{1}{3}]\Gamma[\frac{7}{6}]}{\sqrt{\pi}} \, . \label{time_blowup}
\ee
We can also see that a solution of the non-linear ODE corresponding to a specific choice of initial condition for $\kappa(\tau_0)$ and $\kappa'(\tau_0)$ is,
\be
\kappa(\tau) = \frac{3}{2 (\tau-\tau_b)^2}\, , \label{eq_blowup_kappa}
\ee
which is the blow-up behaviour for all solutions $\kappa(\tau)$ near the blow-up time $\tau_b$; it is characterized by a critical exponent of 2.

We summarise our observations about the PDE in a cosmological framework as follows, where some of them are discussed extensively in a mathematical study being carried out by some of us \cite{PanShi_scale_invariant, PanShi_stability, PanShi_Hamilton}: 
\begin{itemize}
\item The equation $\partial^2_{\tau}{\pi} = (\partial_{r}\pi)^2$ is unstable and blows up at time $\tau_b$ given by Eq.\ (\ref{time_blowup}). For certain initial conditions which are relevant in cosmology, i.e., when the scalar field and its time derivative vanish initially, this is a local phenomenon, in the sense that the blowup point (at a minimum) does not move during its evolution. 
 \item Assuming a small initial value for the scalar field (as a result small $|\kappa (0)|$), we can see that $|\kappa' (0)|$ is sourced by the gravitational potential and the blowup time depends on $\sim |\kappa' (0)|^{-1/3} \sim |{\partial_r^2 \Phi}(\tau_{\rm ini})|^{-1/3}$; a higher curvature of the initial gravitational potential (or equivalently a higher 
 density) 
leads to a faster instability of the system.
 \item Based on the solution we expect that the minima become more curved in time and finally at $\tau = \tau_b$ the curvature becomes infinite.
It is important to note that the mathematical discussion here was based on considering the particular solution (\ref{eq:rsquared}) that is quadratic in $r$.
However, in our mathematical papers we also study this PDE for a more general gravitational potential form (e.g.,  $\Psi(r) = 1- \cos(r) =\frac{1}{2} \pi^{2} r^{2}-\frac{1}{24} \pi^{4} r^{4}+\ldots$ ) where corrections of higher order than $r^2$ contribute,
\be
\pi( \tau,r)=b(\tau) \frac{r^{2}}{2}+d(\tau) \frac{r^{4}}{4 !}+\ldots
\ee
In that case we find a leading order blow-up behaviour $b(\tau) =  \frac{3/2}{(\tau-\tau_b)^2}$ as discussed above, as well as
\be
d(\tau)=\frac{\text { const. }}{\left(\tau_{b}-\tau\right)^{2 \beta-2}}+ ...
\label{eq_d_tau}
\ee
where $\beta = -1.25 + \sqrt{97}/4 = 1.212\ldots$ is a new critical exponent.
\end{itemize}

 \begin{figure*}
\begin{minipage}{\textwidth}
\vspace*{1cm}
 {\hspace*{-1.5cm}{\includegraphics[scale=0.16, trim=-1cm 2cm 0 0cm]{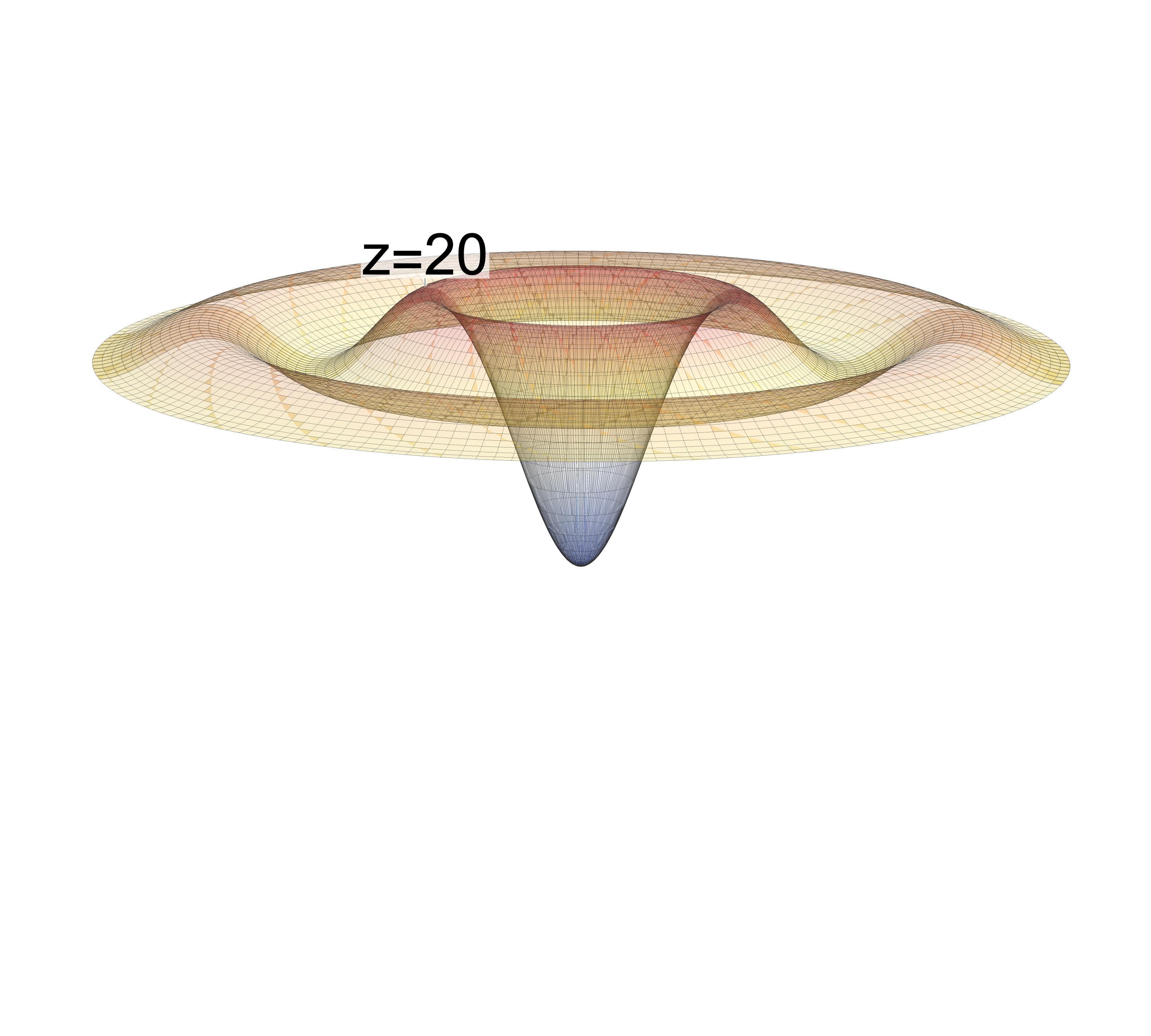} }}
 {\hspace*{-1.5cm}{\includegraphics[scale=0.16, trim=-4cm 3.cm 0 0cm]{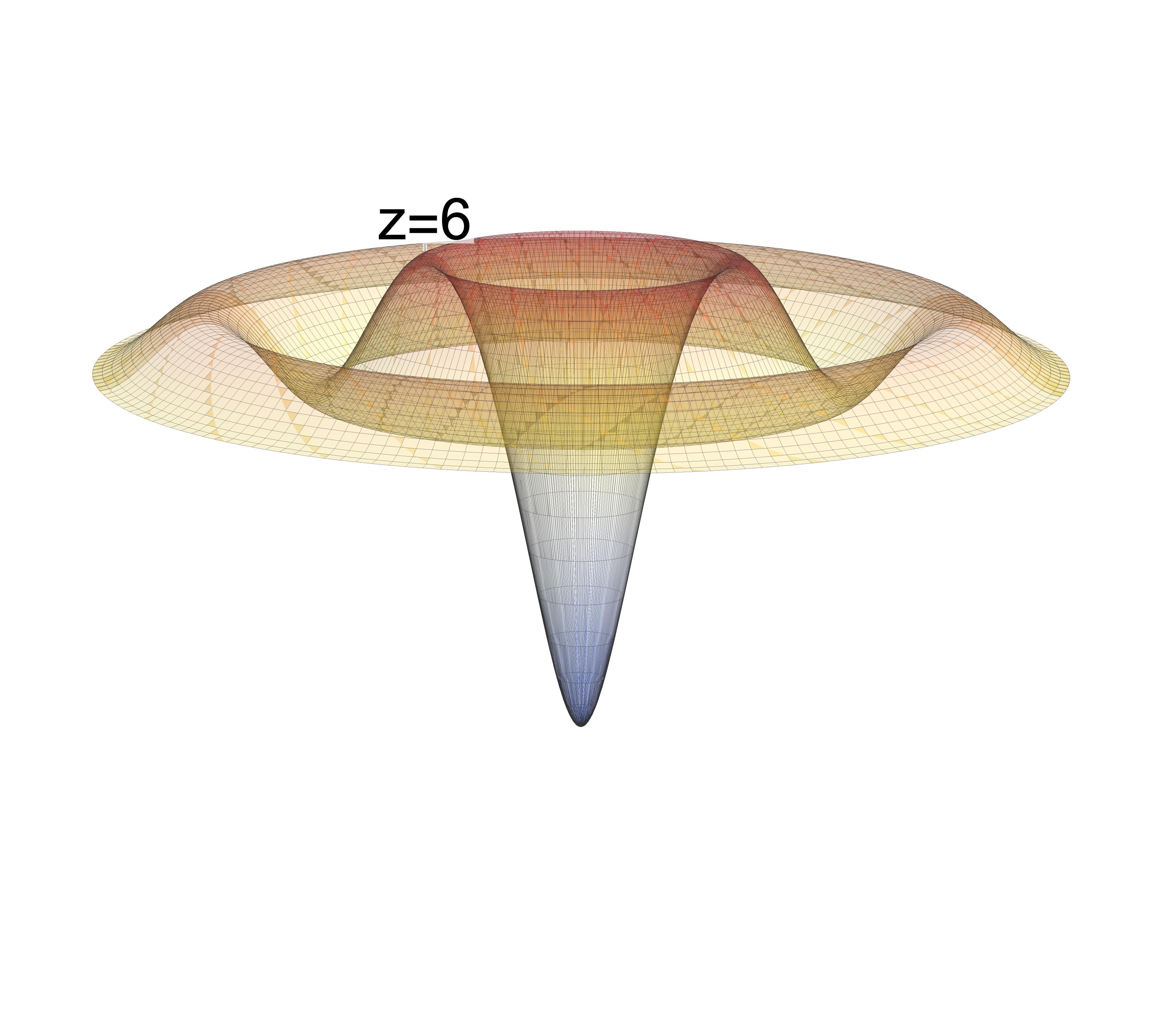} }}
  {\hspace*{-1.5cm}{\includegraphics[scale=0.16, trim=-5cm -1.5cm 0 -1cm]{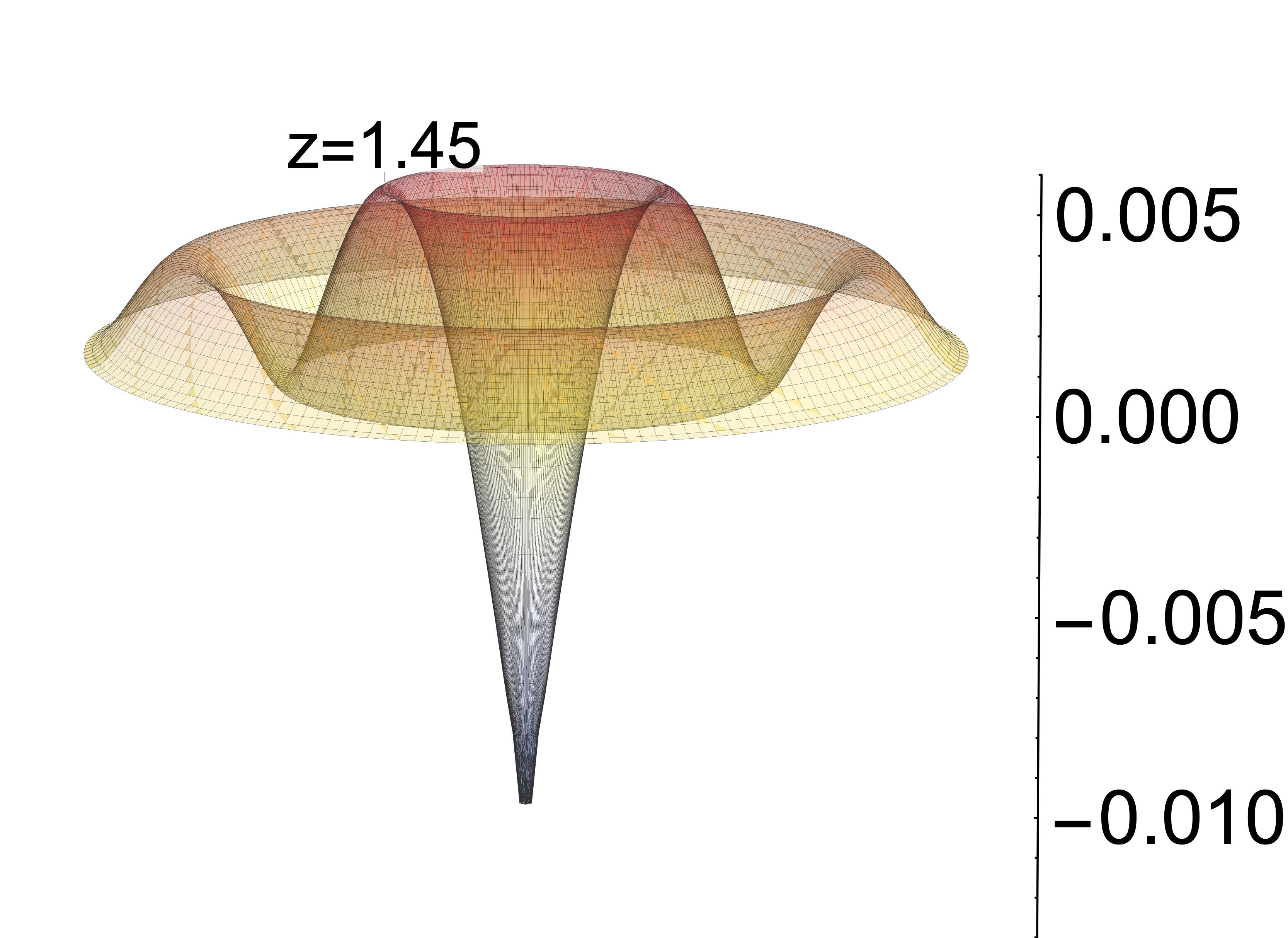} }}\hfill
 \caption{From left to right: The evolution of the scalar field $\pi$ for a spherically symmetric scenario when only the $(\partial_x \pi)^2$ is considered as a non-linear term and all linear terms in Eq.~\ref{full_1Deq} are considered. According to the figure we see a similar behavior compared to the blowup we see in the cosmological $N$-body simulations.}
\label{spherically_sym}
 \end{minipage}
 \end{figure*}

Eq.~\eqref{eq_d_tau} implies that even for non-quadratic initial conditions the instability exists. In Fig.\ \ref{spherically_sym} we show the numerical solution of Eq.~\eqref{1D_PDE} in a 3+1 D spherically symmetric setup. As we can see the minimum of the scalar field becomes sharper in time and develops the instability in a finite time (here at redshift $z=1.45$). Resemblance between our numerical $3+1$D cosmological results with the simplified PDE in Eq.\ (\ref{1D_PDE}) suggests 
that we have correctly identified the source of instability in the full PDE in Eq.~\eqref{full_1Deq}. 

Except for being second order in time the PDE is similar to the Hamilton-Jacobi equation, $\partial_{\tau} \pi = (\nabla \pi)^2$, and even though this does not a priori imply that there should be any relation between the solutions of the two PDEs in our mathematical studies we show that certain aspects of the time evolution of the problem in fact do reflect this analogy.

\section*{Discussion and Conclusions}

This {\it letter} presents a new instability appearing in non-linear PDE's that arise naturally in EFT descriptions of physical problems. We discovered the instability while studying the equations for $k$-essence dark energy in the EFT framework with 3+1D cosmological $N$-body simulations. A mathematical study shows that such non-linear PDE's are unstable and blow up in finite time. This PDE is rich and interesting from a mathematical point of view as it does not seem to fit into any mathematical scheme developed so far.

The potential presence of this instability in the EFT of DE framework for cosmology seems unavoidable as the relevant term, $(\vec{\nabla} \pi)^2$, appears generically for models beyond $\Lambda$CDM. This is almost independent of the physics that the EFT approach is applied to, for example in the EFT of inflation \cite{Cheung:2007st}, similarly to the EFT of DE, this term appears in the second-order equations, and in the EFT of gravity \cite{ArkaniHamed:2003uy} we also expect to have such a term in the equations of motion beyond linear order. Whenever this term is present, and is not balanced by a pressure term $\propto \vec{\nabla}^2 \pi$ with sufficiently large coefficient, then we expect that the solutions cease to exist at finite time, effectively signalling the breakdown of the whole EFT scheme.

Whether the breakdown of the EFT scheme is a sign that such methods cannot be applied to these problems, or whether it points to a fundamental issue with the physical models that it describes, is not yet clear. In the latter case, large classes of models, including low-speed of sound $k$-essence, become unviable and would effectively be ruled out. This could for example be due to shell-crossing in a scalar theory that leads to divergences in the field and the stress-energy tensor. If it is `only' the EFT that fails, then it might be a hint that strong-field effects become important. In this case, and in the cosmological context, it could be that black holes are formed that screen or modify the divergent dynamics. That would be an extremely interesting result as it could help to explain the presence of super-massive black holes in the centers of galaxies.
These questions are the subject of ongoing work. What we can say is that our numerical and analytic studies show that this instability is formed first in the regions with highest density and that the blowup can happen at early times ($z\sim 30$) depending on the density of the center of halos. Moreover, based on numerical studies, this phenomenon is localized, meaning that it does not affect regions that are located well away from the blow-up point.

\paragraph*{Acknowledgements}
We thank Jean-Pierre Eckmann for many interesting discussions and for his comments about the equations and the manuscript. 
This work was supported by a grant from the Swiss National Supercomputing Centre (CSCS) under project ID s1051. We acknowledge funding by the Swiss National Science Foundation.

\bibliography{bibliography}

\end{document}